\documentclass[prd,twocolumn,showpacs,nofootinbib]{revtex4}
\usepackage{times}
\usepackage{natbib}
\usepackage{epsfig}

\newcommand{\bdv}[1]{{\bf #1}}
\newcommand{\up}[1]{{\rm #1}}
\newcommand{\mpc}{{\rm Mpc}}
\newcommand{\gpc}{{\rm Gpc}}

\newcommand{\BB}{\mathcal{B}}   
\newcommand{\gv}{\mathcal{V}}
\newcommand{\CC}{\mathcal{C}}    
\newcommand{\dz}{\delta z}
\newcommand{\nobs}{n_g}
\newcommand{\bear}{\begin{eqnarray}}
\newcommand{\bnobs}{\bar\np}
\newcommand{\zz}{z}      
\newcommand{\DD}{{m_{\dz}}}          
\newcommand{\ax}{\alpha_{\chi}}     
\newcommand{\px}{\varphi_{\chi}}    
\newcommand{\dzg}{\dz_\chi}      
\newcommand{\drg}{\delta\mathcal{R}}  
\newcommand{\rs}{r_s}    
\newcommand{\HH}{\mathcal{H}}   
\newcommand{\ddL}{\delta\mathcal{D}_L}
\newcommand{\kag}{\mathcal{K}}    
\newcommand{\enar}{\end{eqnarray}}
\newcommand{\beeq}{\begin{equation}}
\newcommand{\mdm}{m_\up{dm}}
\newcommand{\ndm}{n_\up{dm}}
\newcommand{\Vang}{\bdv{\hat n}}
\newcommand{\eneq}{\end{equation}}
\newcommand{\mndm}{\langle\ndm\rangle_\Omega}
\newcommand{\oDD}{{{\hat m}_{\dz}}}
\newcommand{\dT}{\delta\tau}        
\newcommand{\gbar}{\bar g}
\newcommand{\vgi}{V}               
\newcommand{\rr}{r}      
\newcommand{\TT}{T}
\newcommand{\mnobs}{\langle\nobs\rangle_\Omega}
\newcommand{\deag}{{\delta e_\chi^\alpha}}      
\newcommand{\pv}{\Psi}           
\newcommand{\dobs}{\delta_g}
\newcommand{\pobs}{P_g}
\newcommand{\coefR}{\mathcal{N}}    
\newcommand{\coefI}{\mathcal{M}}    
\newcommand{\pstd}{P_\up{std}}
\newcommand{\hmpc}{{h^{-1}\mpc}}
\newcommand{\hmpci}{{h\mpc^{-1}}}
\newcommand{\mpci}{{\mpc^{-1}}}
\newcommand{\OM}{\Omega_m}
\newcommand{\OB}{\Omega_b}
\newcommand{\rms}{\sigma_8}
\newcommand{\hgpc}{{h^{-1}\gpc}}
\newcommand{\ct}{\tau}          
\newcommand{\vx}{v_{\chi}}          
\newcommand{\vxo}[1]{v_{\chi #1}}   
\newcommand{\oo}{v}         
\newcommand{\dnu}{\delta\nu}       
\newcommand{\dea}{{\delta e}}      
\newcommand{\cc}{\lambda}   
\newcommand{\dnug}{\delta\nu_\chi}       
\newcommand{\CCG}{\mathcal{G}}   
\newcommand{\nn}{\Vang}  
\newcommand{\dxg}{\delta x_\chi}  
\newcommand{\np}{n_p}
\newcommand{\dLf}{D_L}
\newcommand{\dL}{\mathcal{D}_L}

\begin{document}

\title{General Relativistic Description of the Observed Galaxy Power Spectrum:\\
Do We Understand What We Measure?}

\author{Jaiyul Yoo}
\altaffiliation{jyoo@cfa.harvard.edu} 
\affiliation{Harvard-Smithsonian Center for Astrophysics, Harvard 
University, 60 Garden Street, Cambridge, MA 02138}

\begin{abstract}
We extend the general relativistic description of galaxy clustering
developed in Yoo, Fitzpatrick, and Zaldarriaga (2009). For the first time
we provide a fully general relativistic description of the observed
matter power spectrum and the observed galaxy power spectrum with the
linear bias ansatz. It is significantly different from the standard
Newtonian description on large scales and especially its measurements on large 
scales can be misinterpreted as the detection of the primordial non-Gaussianity
even in the absence thereof. The key difference in the observed galaxy
power spectrum
arises from the real-space
matter fluctuation defined as the matter fluctuation at the hypersurface of
the observed redshift. As opposed to the standard description,
the shape of the observed galaxy power spectrum evolves in
redshift, providing additional cosmological information.
While the systematic errors in the standard Newtonian description are
negligible in the current galaxy surveys at low redshift, 
correct general relativistic
description is essential for understanding the galaxy power spectrum
measurements on large scales in future surveys with redshift depth $z\geq3$.
We discuss ways to improve the detection significance in the
current galaxy surveys and comment on applications of our general
relativistic formalism in future surveys.
\end{abstract}

\pacs{98.80.-k,98.65.-r,98.80.Jk,98.62.Py}

\maketitle

\section{Introduction}
Measurements of the galaxy power spectrum can be used to infer the
shape of the primordial matter power spectrum, providing valuable clues
to understand the initial conditions of early universe. In the linear regime,
galaxy bias ---
the relation between the galaxy and the underlying matter distributions ---
is fairly generic and simple that the shape of the observed galaxy power
spectrum reflects the shape of the matter power spectrum 
\cite{KAISE84}. Therefore, recent theoretical work 
\cite{SMPEET03,MCDON06,CRSC06,YOWEET08}
has focused on interpreting the galaxy power spectrum measurements from 
quasilinear scales to nonlinear scales where the measurement precision 
is highest, yet galaxy bias is significantly affected by the complex nature of
galaxy formation physics.

However,
the past decades have seen a rapid growth in this field, and
large-scale galaxy surveys such as the Sloan Digital Sky Survey 
(SDSS; \cite{YOADET00}) now cover a substantial fraction of the entire sky over
a range of redshift, allowing for measurements of
the galaxy power spectrum 
on sufficiently large scales with unprecedented precision 
\cite{TEEIST06,ROSHET08,REPEET09}.
Recently, \citet{DADOET08} 
showed that the shape of the galaxy power spectrum in the linear regime 
contains a very distinctive feature in the presence of the primordial 
non-Gaussianity. These very large-scale modes of the galaxy power spectrum
are already measured with high signal-to-noise ratios from the SDSS,
putting tight constraints on the amplitude of the
primordial non-Gaussianity \cite{SLHIET08}, 
comparable to the limits from the Wilkinson Microwave Anisotropy Probe (WMAP).
Further theoretical work 
\cite{SLHIET08,AFTO08,MAVE08,MCDON08,FISEZA09,JEKO09,DESE10} shows that
future galaxy surveys can measure the galaxy power spectrum
on large scales with the accuracy enough to detect the small non-Gaussianity
from the simplest single field inflationary models \cite{MALDA03}
and the nonlinear evolution of the Gaussian spectrum \cite{PYCA96}.

The contribution of the primordial non-Gaussianity to the galaxy power
spectrum mainly arises from the gravitational potential and this relativistic
contribution becomes comparable to the contribution of the matter fluctuation
on large scales. However, at this large scale, where the relativistic
effects become dominant, the standard Newtonian description of the galaxy 
power spectrum breaks down and the general relativistic description is 
therefore essential
for understanding the observed galaxy power spectrum and deriving correct 
constraints from the measurements. Furthermore, there exists subtlety in 
theoretical calculations in this regime, where general relativistic effects
are important: Since the general covariance provides a large number of
degrees of freedom, many
theoretical descriptions of observable quantities turn out to have
unphysical gauge freedoms and they fail to correctly describe observable
quantities that we measure on large scales.
Naturally, these systematic errors in those theoretical calculations
are negligible in the Newtonian limit, but they become
substantial on large scales \cite{YOFIZA09}. 

Fully general relativistic descriptions should be constructed by using
observable quantities, rather than theoretically convenient but unobservable
quantities as they are gauge-dependent. The gauge-invariance
is a necessary condition for observable quantities and it should be used
for the consistency check of theoretical calculations of observable quantities.
For example, the observable quantities in the galaxy clustering case 
are the observed redshift, the observed galaxy  position on 
the sky, and the total
number of observed galaxies. \citet*{YOFIZA09} developed a
general relativistic description of galaxy clustering and applied it to the
cross-correlation of large-scale structure with CMB temperature anisotropies,
in which the largest scale modes can be most effectively probed.
On low angular multipoles, the correct relativistic prediction of the
cross-correlation is larger
by about a factor two than the standard Newtonian prediction, alleviating
the discrepancy between the theoretical prediction and the anomalously large 
signal measured from the SDSS and the WMAP data (e.g., see \cite{HOHIET08}).

Here we compute the observed galaxy power spectrum, accounting for all the
general relativistic effects but we only focus on the linear regime,
where the general relativistic effect is substantial.
The observed galaxy power spectrum shows
significant difference on large scales, compared to the standard description,
i.e., the biased matter power spectrum in the synchronous gauge with the 
redshift-space distortion effect \cite{KAISE87}. The correct ``real-space''
matter power spectrum is anisotropic and the shape of the observed
galaxy power spectrum changes with redshift. Especially,
its measurements on large scales can result in a false detection of 
the primordial non-Gaussianity even in the absence thereof.

The organization of this paper is as follows. In Sec.~\ref{sec:mat}
we compute the real-space matter power spectrum accounting
for the relativistic effects and we discuss its stark difference 
from the matter power spectrum in the synchronous gauge. 
We extend the calculation of the real-space 
matter power spectrum to computing the full observed galaxy power spectrum
in Sec.~\ref{ssec:ops} and we forecast the detectability of the departure
of the observed galaxy power spectrum from the standard prediction in the
current and future galaxy surveys in Sec.~\ref{ssec:fore}.
We summarize our new findings and discuss the implication of our 
results for the proposed future surveys in Sec.~\ref{sec:discussion}.

The detailed and technical calculations of our results are presented in
two Appendices. In Appendix~\ref{app:flrw} we summarize our notation
convention and discuss the gauge issues associated with metric representations.
Calculations of
observable quantities are presented in Appendix~\ref{app:gi} with 
particular emphasis on the gauge-invariance of each equation.
In computing our results, we assume there is no vector or tensor mode,
while we present general formulas with vector and tensor modes.
We use the Boltzmann code CMBFast \citep{SEZA96} to obtain the transfer
functions of perturbation variables.

\section{Real-Space Matter power spectrum}
\label{sec:mat}
The general relativistic description of the observed galaxy number density
$\nobs$ is derived in \cite{YOFIZA09} and we rearrange the expression
in terms of gauge-invariant variables defined in Appendix~\ref{app:flrw} as
\bear
\label{eq:full}
\nobs&=&\bnobs(\zz)\bigg[1+
b~\DD+\ax+2~\px+V-C_{\alpha\beta}~e^\alpha e^\beta   \\
&&+3~\dzg+2~{\drg\over\rs}
-H{\partial\over\partial\zz}\left({\dzg\over\HH}\right)
-5p~\ddL-2~\kag\bigg]~, \nonumber
\enar
where $\bnobs$ is the mean galaxy number density in a homogeneous universe
and each term in
the square bracket represents contributions from the volume distortion and
the source effect (see Appendix~\ref{app:gi}).

In this section we compute the contribution of the matter fluctuation 
$\DD=\delta_m-3~\dz$ to the observed galaxy power spectrum.
While the matter fluctuation itself
is not directly observable, computing
the matter power spectrum $P_\DD(\bdv{k})$ 
merits close investigation, since $\DD$ is
boosted by a galaxy bias factor~$b$ separating it from the rest of
the contributions in Eq.~(\ref{eq:full}) and its simple but
gauge-invariant structure provides a guideline for computing the full
observed galaxy power spectrum using Eq.~(\ref{eq:full}).
More importantly, $P_\DD(\bdv{k})$
is the dominant contribution to the observed galaxy power spectrum
on all scales.

\begin{figure}[t]
\centerline{\psfig{file=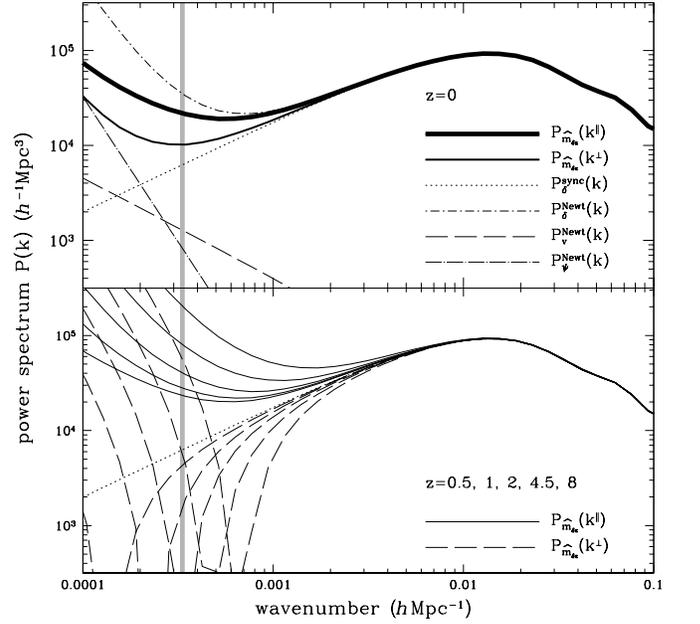, width=3.4in}}
\caption{Matter power spectrum $P_\oDD(k,\mu_k)$ at various redshifts.
Upper panel:
Solid lines represent the matter power spectrum at $z=0$,
computed by using Eq.~(\ref{eq:realP}) 
along the line-of-sight direction ($\mu_k=1$; thick) and along the
transverse direction ($\mu_k=0$; thin).
For reference, various lines indicated in the legend
show power spectra of perturbation variables
in the conformal Newtonian gauge and the synchronous gauge.
Bottom panel: matter power spectrum $P_\oDD(k,\mu_k)$ at $z>0$, 
but with its amplitude normalized to match $P_\oDD(k,\mu_k)$ at $z=0$.
Solid and dashed lines represent $P_\oDD(k,\mu_k)$ with $\mu_k=1$ and
$\mu_k=0$, respectively.
The horizon scale at $z=0$ is shown as a vertical line.}
\label{fig:realp}
\end{figure}

Therefore, we consider the matter fluctuation at the hypersurface defined by
the observed redshift~$\zz$
\beeq
\mdm\ndm(\zz,\Vang)=\bar\rho_m(t)(1+\delta_m)=\bar\rho_m(\zz)(1+\DD)~,
\label{eq:mm}
\eneq
where $\Vang$ is the observed position on the sky,
$\dz$ is the lapse in the observed redshift (see Appendix~\ref{app:gi})
and we assume $\mdm\equiv1$ hereafter.
Of particular importance is the distinction between $\delta_m$ and $\DD$.
The former~$\delta_m$, commonly referred to as the matter fluctuation,
is a {\it gauge-dependent} quantity and its meaning is ambiguous because
the time slicing is unspecified. For example, the same matter density $\rho_m$
can result in different values of~$\delta_m$ due to the change in the mean
matter density $\bar\rho_m(t)$ in different coordinate systems (note that
$\bar\rho_m(t)$ is just a function of coordinate time~$t$).
However, the latter~$\DD$, also we call the matter fluctuation (at the
observed redshift),
is gauge-invariant and unambiguously defined,
since the time slicing is set by the observed redshift~$\zz$, rather than
by an arbitrary choice of coordinate systems as for $\delta_m$.
This perspective provides
a theoretical ground for what ought to be the correct quantity
for the so-called ``real-space'' matter power spectrum 
$P_\DD(\bdv{k})$.\footnote{There are two widely known gauge-invariant
variables for the matter fluctuation $\epsilon_m$ and $\epsilon_g$ defined in
\citet{BARDE80}, where $\epsilon_m=\delta_m+3(1+w)\HH(v+k\beta)/k$ 
and $\epsilon_g=\delta_m+3(1+w)\HH(\beta+\gamma')$
(see Appendix~\ref{app:flrw} for our notation convention). 
They describe the matter fluctuation in the
matter rest frame ($v+k\beta=0$) 
and in the zero shear frame ($\chi=0$), respectively ($\epsilon_m$ and
$\epsilon_g$ correspond to $\delta_v$ and $\delta_\chi$ in our notation).
However, in practice there is no theoretical
reason to prefer one gauge condition to other conditions.
The time slicing in observation is set by the (gauge-invariant)
observed redshift~$\zz$, and hence the matter fluctuation $\DD$ in this
frame correctly represents the ``observed'' matter fluctuation
(note that $\DD=\delta_m$ in the uniform-redshift gauge $\dz=0$).
Furthermore, all the above gauge-invariant variables
$\epsilon_m$, $\epsilon_g$, and $\DD$ are different on large scales
(see Fig.~\ref{fig:realp}),
and the gauge-invariance is not a sufficient condition for observable
quantities.}

In computing the matter power spectrum, we need to compute the observed
mean $\mndm$ and the observed matter fluctuation 
$\oDD\equiv\ndm/\mndm-1$.\footnote{For an observed
number density field $n(\zz,\Vang)$ at
the observed redshift~$\zz$, the 
observed mean $\langle n\rangle_\Omega$ is obtained by averaging $n$ 
over all angle~$\Vang$ within the survey area~$\Omega$,
i.e., averaging over the hypersurface
of simultaneity set by the observed redshift~$\zz$. However, the usual
ensemble average, also commonly denoted as $\langle n\rangle$, is obtained by
averaging $n$ over the hypersurface set by the coordinate time~$t$, and hence
it is gauge-dependent. The observed mean and the theoretical mean (ensemble
average) should be carefully distinguished.} 
The need to distinguish $\oDD$ from $\DD$ arises from the ambiguity of 
quantities at origin in~$\DD$. The full expression of~$\DD$ is
(see Appendix~\ref{app:gi})
\bear
\label{eq:fulldz}
\DD&=&(\delta_m+3~H\chi)-3~(\HH~\dT+H\chi)_o-3~\bigg[\vgi-\alpha_\chi\bigg]_o^s 
\nonumber \\
&+&3\int_0^{\rs}d\rr\bigg[(\ax-\px)'-(\pv_{\alpha|\beta}
+C'_{\alpha\beta})~e^\alpha e^\beta\bigg]~, 
\enar
and it contains quantities evaluated at origin such as
the time lapse~$\dT_o$ and
so on. However, as noted in \citet*{YOFIZA09}, these quantities are
nuisance in observation due to their independence of the observed 
angle~$\Vang$: They are absorbed to the observed mean $\mndm$ and subtracted
from $\ndm$ to give $\oDD$ without the ambiguities at origin
(note that in observation $\oDD$ is derived from the observed number
density $\ndm$).

Though implicit in most theoretical calculations, we explicitly account for
this observational procedure. The observed mean number density is therefore
\bear
\label{eq:correct}
\hspace{-5pt}
\mndm&=&\int_\Omega d^2\Vang~\ndm(\zz,\Vang)\bigg/\int_\Omega d^2\Vang \\
\hspace{-5pt}
&\simeq&\bar\rho_m(z)\bigg[1-3~(\HH_o~\dT_o+H_o\chi_o)+3~(\vgi-\ax)_o
\bigg]~, \nonumber
\enar
where $\vgi=\vgi_\alpha e^\alpha$. 
In the last equality, we assumed that
perturbation variables that vary with the observed angle~$\Vang$
would vanish when averaged over angle within the survey area~$\Omega$.
However, this assumption is valid only if the survey area is infinite:
Modes of wavelength larger than the survey scale would not average out
but set a constant floor in the observed mean. Collectively,
Eq.~(\ref{eq:correct}) may have additional constant. However,
since this additional constant is also independent of observed 
position~$\Vang$ and gauge-invariant,
it has no impact on the power spectrum computation, and we can safely
assume that there is no further contribution in Eq.~(\ref{eq:correct}).
Also, note the subtle difference 
$\mndm\neq\bar\rho_m(\zz)\neq\langle\ndm\rangle=\bar\rho_m(t)$, but
their difference is at the level of metric perturbations.

Therefore, the observed matter fluctuation can be written as 
\bear
\label{eq:mreal}
\hspace{-10pt}
\oDD&\equiv&\ndm/\mndm-1=
(\delta_m+3H\chi)+3~(\alpha_\chi-\vgi)  \nonumber \\
&+&3\int_0^{\rs}d\rr\bigg[(\alpha_\chi-\varphi_\chi)'-(\pv_{\alpha|\beta}
+C'_{\alpha\beta})~e^\alpha e^\beta\bigg]~.
\enar
Now we can make a gauge choice and compute the matter power spectrum
$P_{\oDD}(\bdv{k})$.\footnote{Choosing a gauge condition per se has nothing 
unphysical nor incorrect. In fact, it is desirable for numerical convenience.
The key point is that one has to be careful in relating theoretical 
predictions to observable quantities, for which the gauge-invariance is a 
necessary condition. However, by choosing a gauge condition before deriving
equations, one loses an explicit
way to check the gauge-invariance of theoretical
predictions, and the predictions become gauge-dependent when they involve
unobservable quantities.}
By inspecting Eq.~(\ref{eq:mreal}), it proves 
most convenient to compute the matter power spectrum in the conformal Newtonian
gauge ($\chi=0$),
\beeq
\label{eq:mrealN}
\oDD=\delta_m^N+3~\psi-3~V+3\int_0^{\rs}d\rr~\bigg[(\psi-\phi)'
-H^{T'}_{\alpha\beta}e^\alpha e^\beta\bigg]~,
\eneq
where $\ax=\alpha~(\equiv\psi)$, $\px=\varphi~(\equiv\phi)$ in the conformal
Newtonian gauge.
We emphasize that $\oDD$ computed by using Eq.~(\ref{eq:mrealN}) is
gauge-invariant, i.e., its value is identical to those computed 
by using Eq.~(\ref{eq:mreal}) with
any other choice of gauge conditions.
Finally, the matter power spectrum can be computed as
\bear
\label{eq:realP}
&&\hspace{-15pt}
P_{\oDD}(\bdv{k})=P_\phi(k)~\TT_\oDD(\bdv{k},\zz)~\TT^*_\oDD(\bdv{k},\zz) \\
&&\hspace{-5pt}
\simeq P^N_m(k)+9~P_\psi(k)+6~P_{m\psi}(k)+9~\mu_k^2~P_v(k)~,
\nonumber 
\enar
where $P_\phi(k)$ is the power spectrum of the 
primordial curvature perturbations
and $\TT_\oDD$ is the transfer function for~$\oDD$ (we will use
$\TT_X$ to refer to the transfer function for the perturbation variable~$X$).
We ignored the complication due to survey geometry, and the
redshift dependence of each power spectrum term in the second line is
suppressed for notational simplicity. Since projected
quantities such as the integrated Sachs-Wolfe effect in Eq.~(\ref{eq:mrealN})
are intrinsically angular quantities, they are nearly independent of
the line-of-sight mode~$k^\parallel$ due to the Limber cancellation of
the small scale modes, and we have ignored these contributions in computing
Eq.~(\ref{eq:realP}).

Solid lines in the upper panel of Fig.~\ref{fig:realp} 
represent the matter power spectrum
$P_\oDD(k,\mu_k)$ along the line-of-sight and the transverse directions.
With the time slicing set by the observed redshift~$\zz$, 
the ``real-space'' matter power spectrum $P_\oDD(\bdv{k})$
is no longer isotropic,
and it is neither the matter power spectrum $P^S_\delta(k)$ in the
synchronous gauge (dotted), nor $P^N_\delta(k)$ in the conformal Newtonian gauge
(dot-dashed). Since the synchronous gauge corresponds to the dark matter 
rest frame, $P^S_\delta(k)$ would be what we measure if we knew their positions
without measuring redshift~$\zz$. However, the (coordinate)
positions in the synchronous
gauge are also gauge-dependent quantities, and we need the observed
redshift~$\zz$ to define the time slicing and their observed positions.
On small scales $k\gg10\HH$, the matter power spectrum $P_\oDD(k,\mu_k)$
becomes virtually isotropic, and the difference in $P_\delta(k)$ with various
gauge conditions vanishes compared to the correct matter power spectrum
$P_\oDD(k,\mu_k)$.

The bottom panel shows the matter power spectrum $P_\oDD(k,\mu_k)$ at
high redshifts $z=0.5$, 1, 2, 4.5, 8 along the line-of-sight direction
(solid) and along the transverse direction (dashed).
Compared to $P_\oDD(k^\perp)$ at $z=0$,
significant difference is immediately noticeable along the transverse
direction, where the matter power spectrum has the zero-crossing scale
at each redshift. Along the transverse direction, the matter fluctuation
is $\oDD\simeq\delta_m^N+3~\psi$. Since the matter fluctuation~$\delta_m^N$ 
decreases with redshift but the gravitational potential $\psi~(<0)$ remains
nearly constant, the matter fluctuation $\oDD$ becomes zero and changes its
sign on large scales, and this zero-crossing scale becomes smaller at
higher redshift. Physically, the magnitude of the lapse~$\dz$ in the
observed redshift is larger than the density fluctuation $\delta_m$
on large scales,
such that the change in the mean matter density $\bar\rho_m(t)$
overwhelms the density
fluctuation and hence $\oDD$~appears to be underdense in observation.
The matter power spectrum $P_\oDD(k^\parallel)$ along the
line-of-sight direction maintains the similar trend at $z=0$ shown in
the upper panel. However, compared to the matter power spectrum $P_\delta^S(k)$ 
in the synchronous gauge (dotted), 
the turn-off scale of $P_\oDD(k,\mu_k)$ becomes smaller,
as the horizon size decreases with redshift.

\section{Observed Galaxy Power Spectrum}
\label{sec:ops}
\subsection{Theoretical Prediction}
\label{ssec:ops}

Having computed the real-space matter power spectrum in Sec.~\ref{sec:mat}, 
we are now in a position to
tackle the full complexity of the observed galaxy number density $\nobs$ 
in Eq.~(\ref{eq:full}). Following the same procedure, we compute the observed
mean galaxy number density by considering contributions independent of
the observed galaxy position,
\beeq
\label{eq:corrmn}
\mnobs=\bnobs\bigg[1+b~\DD_o+3~{\dzg}_o+2~{{\drg}_o\over\rs}
-5p~{\ddL}_o-2~{\kag}_o\bigg]~,
\eneq
where the surface terms in each component are
\bear
&&{\dzg}_o=(\HH_o\dT_o+H_o\chi_o)-(\vgi-\ax)_o~, \\
&&\DD_o=-3~{\dzg}_o~, \nonumber \\
&&{\drg}_o=(\chi_o+\dT_o)-{{\dzg}_o\over\HH}~, \nonumber \\
&&{\ddL}_o=(\HH_o\dT_o+H_o\chi_o)+\left({\chi_o+\dT_o\over\rs}\right)
-{{\dzg}_o\over\HH\rs}~, \nonumber \\
&&{\kag}_o=e_\alpha~(\deag+\pv^\alpha+2~C^\alpha_\beta~e^\beta)_o~. \nonumber
\enar
Therefore, the observed galaxy fluctuation can be obtained by subtracting
the observed mean as
\bear
\label{eq:final}
\hspace{-10pt}
\dobs&=&\nobs/\mnobs-1=b~(\DD-\DD_o)+\ax+2~\px   \\
&&\hspace{-10pt}
+V-C_{\alpha\beta}~e^\alpha e^\beta
+3~(\dzg-{\dzg}_o)+2~\left({\drg-{\drg}_o\over\rs}\right) \nonumber \\
&&\hspace{-10pt}
-H{\partial\over\partial\zz}\left({\dzg\over\HH}\right)
-5p~(\ddL-{\ddL}_o)-2~(\kag-{\kag}_o)~. \nonumber
\enar
This equation maintains the gauge-invariant structure, yet lacks any ambiguity
at origin.

As discussed in Sec.~\ref{sec:mat}, we make
simplifications to facilitate computing the observed galaxy power spectrum
$\pobs(k,\mu_k)$ by adopting a simple survey geometry,
ignoring contributions from the projected quantities,
and choosing the conformal Newtonian gauge ($\chi=0$). The observed galaxy
power spectrum is therefore
\beeq
\pobs(\bdv{k})= P_\phi(k)~\TT_g(\bdv{k},\zz)~\TT^*_g(\bdv{k},\zz)~,
\label{eq:ppp}
\eneq
with the transfer function
\bear
\TT_g&=&b~(\TT^N_m+3~\TT_\psi-3~\TT_\vgi)+\TT_\psi+2~\TT_\phi+\TT_\vgi \nonumber\\
&&+3~(\TT_\vgi-\TT_\psi)+2~\left({\TT_\psi-\TT_\vgi\over\HH\rs}\right)
-H{\partial\over\partial\zz}\left({\TT_\vgi-\TT_\psi\over\HH}\right) \nonumber \\
&&-5p~\left(\TT_\vgi-\TT_\psi+{\TT_\psi-\TT_\vgi\over\HH\rs}\right) \nonumber \\
&=&b~\TT^N_m+\coefR\TT_\psi-{\TT_{\phi'}\over\HH}
-\mu_k^2~{k\TT_v\over\HH}+i\mu_k\coefI\TT_v~,
\enar
where two coefficients in the transfer function are
\bear
&&\coefR=3~b-1+{1+z\over H}{dH\over dz}+(5p-2)-\left({5p-2\over\HH\rs}\right)~,
\nonumber \\
&&\coefI=3~b+{1+z\over H}{dH\over dz}+(5p-2)-\left({5p-2\over\HH\rs}\right)~.
\enar
Finally, the observed galaxy power spectrum in Eq.~(\ref{eq:ppp}) is 
\bear
\label{eq:fullpobs}
\pobs(k,\mu_k)&=& b^2P^N_m(k)+\coefR^2P_\psi(k)+2~b\coefR P_{m\psi}(k) 
\nonumber \\
&+&\mu_k^2\left[\coefI^2P_v(k)-{2~b~k\over\HH}P_{mv}(k)
-{2~\coefR~k\over\HH}P_{\psi v}(k)\right] \nonumber \\
&+&\mu_k^4~{k^2\over\HH^2}P_v(k)~, 
\enar
where we ignored the contribution from $\TT_{\phi'}/\HH$, which vanishes
quickly at $z>0$. 
With  our fiducial cosmology these coefficients range
$\coefR-3~b=-0.5\sim0.5$ and $\coefI-3~b=0.8\sim1.5$ for the luminosity
function slope $5p=2$ of the source galaxy sample, where the source effect
is minimized.\footnote{The source effect cannot be exactly cancelled 
on all scales, even for the case $5p=2$, because the fluctuation $\ddL$ in the 
luminosity distance reduces to the standard convergence $\kappa$
only in the Newtonian limit, i.e., $5p~\ddL+2~\kag\neq0$ on large scales,
even when $5p=2$.}
The observed galaxy power spectrum $\pobs(k,\mu_k)$ is therefore approximately
the biased matter power spectrum $b^2P_\oDD(k,\mu_k)$ with the 
standard redshift-space distortion factor, in which case two coefficients
are exactly $\coefR=3~b$, $\coefI=3~b$ (when $5p=2$ is assumed).

In comparison, the observed galaxy power spectrum and its transfer function
in the standard method are
\bear
\label{eq:stdpok}
\pstd(k,\mu_k)&=&P_\phi(k)~\TT_\up{std}(\bdv{k},z)~\TT^*_\up{std}(\bdv{k},z)\\
&=&b^2P_\delta^S(k)-\mu_k^2~{2~b~k\over\HH}P_{\delta v}(k)
+\mu_k^4~{k^2\over\HH^2}P_v(k)~, \nonumber
\enar
and
\beeq
\TT_\up{std}=b~\TT^S_m-\mu_k^2~{k\TT_v\over\HH}~,
\eneq
respectively. The standard transfer function $\TT_\up{std}$ is composed
of two transfer functions ($\TT^S_m$ and $\TT_v$) in two different gauge
conditions, one in the synchronous gauge and one in the conformal Newtonian
gauge.

\begin{figure}[t]
\centerline{\psfig{file=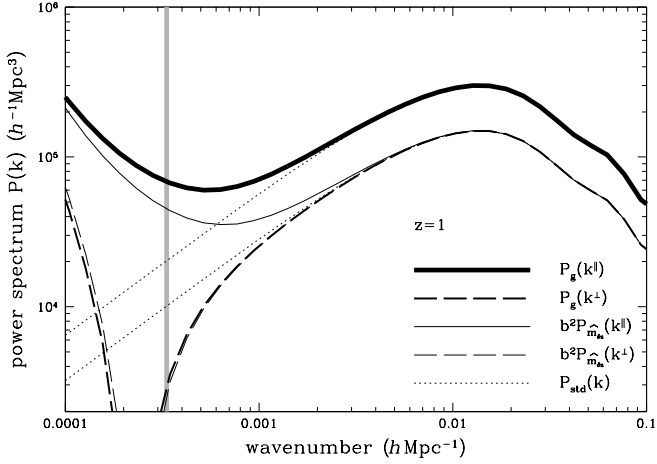, width=3.4in}}
\caption{Observed galaxy power spectrum $\pobs(k,\mu_k)$ at $z=1$.
Thick lines represent the observed galaxy power spectrum, computed
by using the full general relativistic description in
Eq.~(\ref{eq:fullpobs}) along the line-of-sight direction
($\mu_k=1$; solid) and along the transverse direction ($\mu_k=0$; dashed).
Dotted lines show the observed galaxy power spectrum using the standard method 
in Eq.~(\ref{eq:stdpok}).
For comparison, the matter power spectrum $b^2P_\oDD(k,\mu_k)$ is shown as
thin lines with the galaxy bias factor $b=2$.}
\label{fig:full}
\end{figure}

Figure~\ref{fig:full} shows the observed galaxy power
spectrum $\pobs(k,\mu_k)$ at $z=1$, computed by using Eq.~(\ref{eq:fullpobs}).
As we noted in Sec.~\ref{sec:mat}, the matter power spectrum $P_\oDD(k,\mu_k)$
is the dominant contribution to the observed galaxy power spectrum
on all scales, and
the redshift-space distortion from the volume distortion  enhances the power.
Along the transverse direction (thick dashed), 
the observed galaxy power spectrum $\pobs(k^\perp)$
largely traces the biased real-space matter power spectrum $b^2P_\oDD(k^\perp)$ 
(thin dashed) as  the standard description of the observed galaxy power 
spectrum (lower dotted) is the biased matter power spectrum in the synchronous
gauge $\pstd(k^\perp)=b^2P^S_\delta(k)$. However, on large scales,
the standard description overestimates
the observed galaxy power spectrum along the transverse
direction due to the contribution of~$\dz$ in $P_\oDD(k^\perp)$.
Similarly, the observed galaxy power 
spectrum $\pobs(k^\parallel)$ along the line-of-sight direction (thick solid)
significantly deviates from the standard description
$\pstd(k^\parallel)=(b+f)^2P^S_\delta(k)$ (upper dotted) on large 
scales, where $f$ is the growth rate of structure. The deviation 
arises mainly from the difference between $b^2P_\oDD(k^\parallel)$ and 
$b^2P^S_\delta(k)$, but there exists small contribution
from the volume distortion in addition to the standard
redshift-space distortion effect.

\begin{figure}[t]
\centerline{\psfig{file=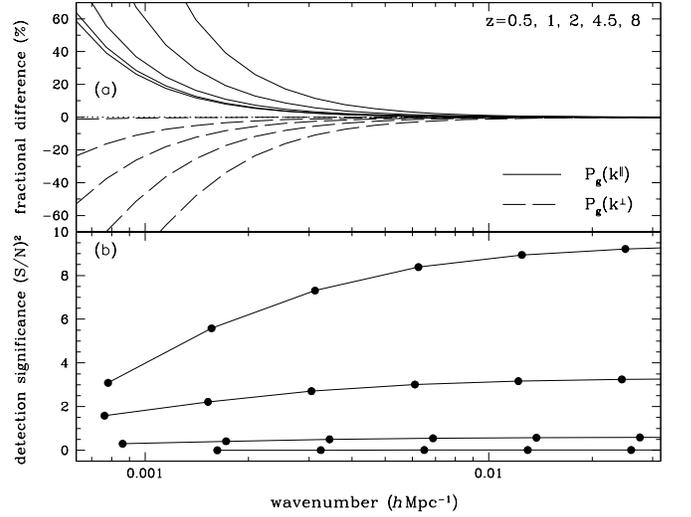, width=3.4in}}
\caption{Systematic errors in theoretical modeling of the observed galaxy
power spectrum. Upper panel: fractional difference
of $\pobs(k,\mu_k)$ at various redshift slices,
compared to the standard description
$\pstd(k,\mu_k)$ in Eq.~(\ref{eq:stdpok})
along the line-of-sight direction (solid) and along the
transverse direction (dashed). 
Bottom panel: detection significance of the departure from the standard
description in a cosmic-variance limited survey as a function of maximum
wavenumber. The survey volume is divided
into four spherical shells with redshift range
$z=0\sim1$ (bottom solid), $1\sim3$, $3\sim6$, and $6\sim10$ (top solid),
and the signal is computed at the mean redshift of each shell.}
\label{fig:dev}
\end{figure}

Figure~\ref{fig:dev}a compares
the correct galaxy power spectrum $\pobs(k,\mu_k)$ in 
Eq.~(\ref{eq:fullpobs}) with respect to 
the standard description $\pstd(k,\mu_k)$ in Eq.~(\ref{eq:stdpok}) 
at various redshift slices.  Solid and dashed lines represent the fractional
difference $\Delta P_g/\pstd=\pobs/\pstd-1$ along the line-of-sight direction
($\mu_k=1$; solid) and along the transverse direction ($\mu_k=0$; dashed),
respectively. On large scales $k\leq0.01\hmpci$,
the standard description underestimates the observed galaxy
power spectrum along the line-of-sight direction and overestimates
along the transverse direction. 
At higher redshift, the fractional difference in 
the observed galaxy power spectrum at a fixed wavenumber
is progressively amplified,
because the relativistic effects increase as the horizon size
becomes smaller. A few percent deviation is already present at $k=0.01\hmpci$
at $z\geq4.5$.

\subsection{Statistical uncertainties on power spectrum measurements}
\label{ssec:fore}

Of the utmost importance is the detectability of the systematic errors
in theoretical modeling of the observed galaxy power spectrum 
$\pobs(k,\mu_k)$ on large scales.
We consider statistical uncertainties on measurements of the two-dimensional 
anisotropic power spectrum by using a simple mode counting approximation.
Since Fourier modes are effectively independent
of each other, the statistical uncertainty decreases with 
$1/\sqrt{N_k}$, where $N_k$ is the number of Fourier modes available given
a survey volume $V_s$. 
However, sampling of individual galaxies contributes the shot
noise errors $1/\mnobs$ to the signal $\pobs(k,\mu_k)$,
and the number of independent Fourier
modes is in fact a half of $N_k$ as $\dobs(\bdv{k})$ represents a real
quantity in configuration space. Therefore, the statistical uncertainty 
of each Fourier mode is \cite{FEKAPE94,TEGMA97}
\beeq
\left(\sigma_p\over\pobs\right)^2={2\over V_kV_\up{eff}}~,
\eneq
and the detection significance of the departure in the observed galaxy power
spectrum from the standard method prediction is 
\beeq
\left({S\over N}\right)^2=\left({\Delta P_g\over\sigma_p}\right)^2
=\left(\Delta P_g\over\pstd\right)^2\left({V_kV_\up{eff}\over2}\right)~,
\eneq
where $\Delta P_g=\pobs-\pstd$, 
\beeq
V_{\bdv{k}}={2\pi\Delta(k^\perp)^2\Delta k^\parallel\over(2\pi)^3}~,
\eneq
is the Fourier volume accounting for the positive and negative $k^\parallel$,
and the effective survey volume is
\beeq
V_\up{eff}=\int dV_s\left[{\mnobs\pobs\over1+\mnobs\pobs}\right]^2~.
\label{eq:veff}
\eneq

For definiteness we consider galaxy samples measured
from a large-scale survey with full sky coverage
(cosmic-variance limited) as a function of survey redshift depth.
We choose the galaxy number density
$\mnobs=10^{-4}(\hmpc)^{-3}$, similar to what the current large-scale
surveys aim to measure (e.g., \cite{EIANET01}).
Since the shot noise contribution is small $\mnobs\pobs\geq1$ 
on scales of our interest $k\leq0.01\hmpci$,
the effective volume in Eq.~(\ref{eq:veff}) is therefore
identical to the survey volume $V_\up{eff}\simeq V_s$.
Here we assume a flat $\Lambda$CDM
universe with the matter density $\OM=0.24$,
the baryon density $\OB=0.042$, the Hubble constant $h=0.73$,
the spectral index $n_s=0.954$, the optical depth to
the last scattering surface $s=0.09$, and the primordial
curvature perturbation amplitude $\Delta_\phi^2=2.38\times10^{-9}$
at $k=0.05~\mpci$ ($\rms=0.81$). 

The significance $(S/N)^2$ of the systematic errors
as a function of maximum wavenumber
is shown in Fig.~\ref{fig:dev}b, where we divide the survey volume into four
spherical shells, covering the full sky with redshift range 
$z=0\sim1$, $1\sim3$, $3\sim6$, and $6\sim10$. At the lowest redshift bin
($z=0\sim1$), the survey volume is $V_s=58~(\hgpc)^3$ and the minimum
wavenumber is $k_\up{min}=\Delta k=2\pi/V_s^{1/3}=0.0016\hmpci$, at which
the fractional difference is 8\% along the line-of-sight direction 
(solid) and is close to zero along the transverse direction (dashed)
as shown in Fig.~\ref{fig:dev}a. With only a handful of modes at $k_\up{min}$,
the systematic errors in theoretical modeling 
of the observed galaxy power spectrum is negligible at $z\leq1$
(bottom solid line in Fig.~\ref{fig:dev}b).
Similar conclusion can be derived for the survey volume at $z=1\sim3$ 
(second solid line from the bottom).

However, at the higher redshift bins (two upper solid lines),
there exist two critical differences in assessing
the significance of the systematic errors: 
Larger survey volume and smaller horizon size.
About a factor of ten larger volume is available in the higher redshift bins
than in the lowest redshift bin, which reduces the statistical uncertainties
at a fixed~$k$ by $10^{1/2}\simeq3.2$
and provides $k_\up{min}$ smaller by $10^{1/3}\simeq2.2$.
Furthermore, the relativistic effects are larger at a fixed~$k$
due to the smaller horizon size at the higher redshift bins, increasing the
fractional difference $\Delta P_g/\pstd$ at each~$k$.
Consequently, the  significance of the systematic errors
increases dramatically with survey redshift depth.

In conclusion, the volume available at $z\leq3$ is less than the Hubble
volume and hence the standard Newtonian description is statistically
indistinguishable from the general relativistic description of the 
observed galaxy power spectrum at $z\leq3$. However,
the systematic errors in theoretical modeling increase substantially with
redshift and correct general relativistic description is essential 
in understanding the observed galaxy power spectrum at $z\geq3$.

In practice, further complication will be present
for measuring the largest scale
modes  in galaxy surveys, because they are affected by survey
window functions and large contiguous region is required. Long
line-of-sight modes are relatively immune to these difficulties, while
modeling the redshift evolution of galaxy bias may complicate the
theoretical interpretation of the observed galaxy power spectrum
$\pobs(k^\parallel)$ along the line-of-sight
direction. Proper application of our method to observations will require
further investigation with close ties to the specifications of a given
 survey geometry.

\section{DISCUSSION}
\label{sec:discussion}
We have extended the general relativistic description of galaxy
clustering developed in \citet*{YOFIZA09} and we have provided, for the first 
time, the correct general relativistic description of the observed galaxy
power spectrum with particular emphasis on 
the expressions for observable quantities and their gauge-invariance.
In the galaxy clustering case, the observable quantities involve the observed
redshift~$\zz$, the galaxy position~$\Vang$ on the sky, the number $N_\up{tot}$
of observed galaxies, and its apparent luminosity~$L$. These quantities are
different from the quantities defined in a homogeneous and isotropic universe,
which are theoretically more tractable, yet unobservable (gauge-dependent)
quantities. The gauge-invariance is a necessary condition for observable
quantities and the observed galaxy power spectrum should be expressed in terms
of the observable quantities.

There are two key contributions to the observed galaxy power spectrum:
The real-space matter fluctuation and the volume distortion. While the
standard redshift-space distortion effect accounts for most of the 
contributions from the volume distortion, the former requires a careful
definition of its meaning and it exhibits significant difference on large
scales compared to the standard real-space matter fluctuation. Since the
observed redshift~$\zz$ defines the hypersurface of simultaneity in 
observation, the correct description of the real-space matter fluctuation
is the matter fluctuation at the observed redshift,
$\DD=\delta_m-3~\dz$ ($\DD=\delta_m$ in the uniform-redshift
frame $\dz=0$). As it is defined by observable quantities, $\DD$~is
naturally gauge-invariant, in contrast to the usual (gauge-dependent)
matter fluctuation~$\delta_m$, which requires a specification of gauge
conditions or coordinate systems and differs in its value contingent upon
the gauge choice.

As the observed redshift~$\zz$ depends on angle,
the real-space matter power spectrum $P_\DD(\bdv{k})$ is no longer isotropic:
Compared to the matter power spectrum $P^S_\delta(k)$ in the synchronous
gauge, $P_\DD(\bdv{k})$ is enhanced along the line-of-sight direction but
it is suppressed along the transverse direction, because the lapse~$\dz$
in the observed redshift can increase or decrease the mean matter density,
affecting~$\delta_m$ in both direction.  
Consequently, the shapes of the real-space matter power spectrum 
$P_\DD(\bdv{k})$ and the observed galaxy power spectrum vary with redshift
at $k\leq0.01\hmpci$, while the shape of the matter power spectrum
$P_\delta^S(k)$ in the synchronous gauge is independent of redshift.
Therefore, a proper modification is necessary for methods to use
the multipole moments of the observed galaxy power spectrum on large scales
to constrain the growth factor of the matter fluctuation.
Traditionally, measurements of the observed galaxy power spectrum 
at different redshift slices are combined to reduce the statistical 
uncertainties, but with the redshift evolution of the power spectrum shape
on large scales, they can be used to extract additional cosmological 
information.

Recently, it is shown \cite{DADOET08} that the primordial non-Gaussianity in
gravitational potential can give rise to the scale-dependence of 
the observed galaxy power spectrum
on large scales, where galaxy bias is assumed to be linear.
Furthermore, it is guaranteed \cite{AFTO08,MCDON08,CAVEMA08}
 that future galaxy surveys will
detect non-Gaussian signatures on large scales, arising from either primordial
or nonlinear evolution. However, 
its signature on large scales is similar to the rising power of the
real-space matter power spectrum
$P_\DD(\bdv{k})$ in Fig.~\ref{fig:realp} and 
the observed galaxy power spectrum in
Fig.~\ref{fig:full}, since they all originate from the contribution of the
gravitational potential. Therefore, without proper theoretical modeling
of the observed galaxy power spectrum,
its measurements on large scales can be misinterpreted
as the detection of the primordial non-Gaussianity even in the absence thereof,
and its false detection
can be used as the evidence against the simplest single field inflationary 
models. With the full general relativistic description of 
the observed galaxy power spectrum developed here, the effect of the
primordial non-Gaussianity on 
the observed galaxy power spectrum will be investigated in
a separate work.

While the current galaxy surveys
  cover a large fraction of the entire sky, it is still
difficult to directly measure the departure of the observed
galaxy power spectrum from the standard description
 at $z\leq3$ with high significance,
simply due to the limited number of large-scale modes. However,
recent work \cite{SELJA09} suggested a method to eliminate the cosmic variance
errors by comparing two different biased tracers. The scale-dependence of
the ratio of two galaxy power spectra arises from the volume distortion: Unlike
in the standard description of the primordial non-Gaussianity, all biased
traces depend on the same matter fluctuation~$\DD$ and hence there would be
no scale-dependence, if it were not for the volume distortion.
With substantial increase in the signal-to-noise ratio, 
current galaxy surveys may be able to detect the general relativistic
effects in the observed galaxy power spectrum at low redshift.

In future galaxy surveys with larger redshift depth,
correct general relativistic description should be
an essential element in power spectrum analysis, as
significant systematic errors will be present
in the standard description of the observed galaxy power spectrum 
due to smaller horizon size and larger survey volume at high redshift.
Furthermore, our formalism can be easily extended to any alternative theories 
of gravity. Since these theories are identical in the Newtonian limit but
have distinctive relativistic effects, alternative theories of gravity can be
most effectively probed on cosmological scales. Comparison of their theoretical
predictions
to the galaxy power spectrum measurements on large scales
can, therefore, provide a way 
to test general relativity and put constraints on the alternative.

While we considered a spectroscopic galaxy survey here, a majority of 
planned large-scale surveys such as the Dark Energy Survey (DES), 
the Panoramic Survey Telescope \& Rapid Response System (PanSTARRS),
and the Large Synoptic Survey Telescope (LSST)
lack of observationally expensive spectroscopic 
instruments, and instead they rely on photometric redshift measurements.
However, the degradation of redshift measurement precision will have little
impact on the detectability of large-scale modes, since it only results 
in radial smearing of galaxy positions ($\Delta r\leq100\hmpc$ at $z=1$,
and smaller at higher redshift) and the general relativistic effect 
appears on large scales $k\leq0.01\hmpci$.

\acknowledgments
J.Y. acknowledges useful discussions with Daniel Eisenstein,
Eiichiro Komatsu, Uro{\v s} Seljak, and Michael Turner. Especially,
J.Y. thanks Matias Zaldarriaga for his encouragement and critical comments
throughout the completion of this work.
J.~Y. is supported by the Harvard College Observatory under the
Donald~H. Menzel fund.

\vfill

\bibliography{ms.bbl}

\appendix
\section{FLRW metric and notation convention}
\label{app:flrw}
We summarize our notation convention 
for the most general metric in an inhomogeneous 
Friedmann-Lema{\^\i}tre-Robertson-Walker (FLRW) universe and we discuss the
gauge transformation properties and gauge freedoms
of the perturbation variables used in this paper.

The background universe is described by a spatially homogeneous and isotropic
metric with scale factor $a(t)$,
\beeq
\label{eq:bgmetric}
ds^2=g_{ab}~dx^a~dx^b=-dt^2+a^2(t)~\gbar_{\alpha\beta}~dx^\alpha~ dx^\beta~,
\eneq
where $\gbar_{\alpha\beta}$ is the metric tensor for a three-space with a 
constant spatial curvature $K=-H_0^2~(1-\Omega_\up{tot})$. We extensively
use the conformal time~$\ct$, defined as $a(\ct)~d\ct=dt$~.
Throughout the paper, we use the Latin indices for the
spacetime component and the Greek indices for the spatial component, and 
we set the speed of light $c\equiv1$.

Small departure of the metric tensor
in the inhomogeneous universe from the background metric
can be represented as a set of metric perturbations:
\bear
\delta g_{00}&=&-2~a^2~\alpha~, \\
\delta g_{0\alpha}&=&-a^2\BB_\alpha=-a^2(\beta_{,\alpha}+B_\alpha)
 ~, \nonumber \\
\delta g_{\alpha\beta}&=&2~a^2\CC_{\alpha\beta}=
2~a^2~(\varphi~\gbar_{\alpha\beta}
+\gamma_{,\alpha|\beta}+C_{(\alpha|\beta)}
+C_{\alpha\beta})~, \nonumber
\enar
where the vertical bar represents the covariant derivative with respect to
spatial metric $\gbar_{\alpha\beta}$~, and the perturbation classification
from its transformation properties is represented with single or double
indicies for vector-type or tensor-type, respectively.
We adopt the classification scheme introduced in \cite{BARDE80}.

The general covariance in general relativity guarantees that any coordinate
system can be used to describe physical systems and the choice of a coordinate
system bears no physical significance. A coordinate transformation accompanies
a change in the correspondence of the real physical (inhomogeneous) universe to
the homogeneous background universe, known as the gauge transformation.
Here we consider the most general coordinate transformation
\beeq
\tilde x^a=x^a+\xi^a~,
\label{aeq:gt}
\eneq
where $\xi^a=(T,\mathcal{L}^\alpha)$ and $\mathcal{L}^\alpha=L^{,\alpha}+L^\alpha$. 
Under the coordinate transformation, the scalar metric perturbations
transform as
\bear
\tilde\alpha&=&\alpha-T'-\HH~T    ~,\\
\tilde\beta&=&\beta-T+L'         ~,\nonumber \\
\tilde\varphi&=&\varphi-\HH~T    ~, \nonumber \\
\tilde\gamma&=&\gamma-L     ~, \nonumber 
\enar
where the conformal Hubble parameter is $\HH=a'/a=aH$ and 
we denote the conformal time derivative as a prime. 
The vector metric perturbations transform as
\bear
\tilde B_\alpha&=&B_\alpha+L'_\alpha    ~, \\
\tilde C_\alpha&=&C_\alpha-L_\alpha    ~. \nonumber 
\enar
Since tensor harmonics are independent of tensors
that can be constructed from coordinate transformations,
tensor-type perturbations remain unchanged under the gauge transformation.

Based on the gauge transformation
properties, fully gauge-invariant quantities can be constructed as
\bear
\ax&=&\alpha-\chi'/a       ~, \\
\px&=&\varphi-H\chi        ~, \nonumber \\
\vx&=&v+k~\beta-k~\chi/a            ~, \nonumber 
\enar
for scalar perturbations, where $\chi=a~(\beta+\gamma')$ and
 $\tilde\chi=\chi-aT$~.
Our notation is chosen to facilitate the choice of gauge conditions
\cite{HWNO01}, e.g.,
 $\ax=\alpha$, $\px=\varphi$, and $\vx=v$ in the conformal Newtonian
gauge ($\beta=\gamma=0$) or the zero-shear gauge ($\chi=0$).
For vector perturbations, two gauge-invariant quantities are
\bear
\pv_\alpha&=&B_\alpha+C'_\alpha~,  \\
\gv_\alpha&=&v_\alpha-B_\alpha ~, \nonumber
\enar
and we have used $\vgi_\alpha=\vxo{,\alpha}+\gv_\alpha$ in the text.
These gauge-invariant variables correspond to $\Phi_A$, $\Phi_H$, 
$v^{(0)}_s$, $\Psi$, and $v_c$ in Bardeen's notation \cite{BARDE80}. 
In addition to these gauge-invariant variables, one can construct other
gauge-invariant variables, e.g., $\delta_v=\delta+3(1+w)\HH(v+k\beta)/k$,
where $\delta_v=\delta$ in the comoving gauge, and indeed there are
as many possibilities for gauge-invariant variables as for the choice of
gauge condition.

Due to the spatial homogeneity of the background universe, all the
perturbations should be invariant under pure spatial gauge transformations
($T=0$, $\mathcal{L}^\alpha\neq0$).  Therefore, those perturbation variables
$(\beta,\gamma,B_\alpha,C_\alpha)$
that transform with $L$ or $L_\alpha$ carry unphysical gauge freedoms
and they can appear in physical quantities
only through the combinations $\chi$ and $\pv_\alpha$
that are invariant under spatial
gauge transformations \cite{BARDE88}. When the observed mean is subtracted
from the observed number density field
in Secs.~\ref{sec:mat} and~\ref{sec:ops}, these perturbation variables 
$(\beta,\gamma,B_\alpha,C_\alpha)$
may leave unphysical gauge freedoms in the observed fluctuation field
ruining its gauge-invariance. Fortunately, the damage is relatively innocuous,
affecting only the monopole and the dipole at origin. In the paper, we
explicitly resolve this issue by rearranging physical quantities
in terms of spatially gauge-invariant combinations~$\chi$ and~$\pv_\alpha$,
before any operation is taken.

\section{Derivation of the Gauge-Invariant Equations}
\label{app:gi}
Here we derive the gauge-invariant equations used in the power spectrum
analysis. The derivation closely follows our previous work \cite{YOFIZA09},
but we pay particular attention to the gauge-invariance of perturbation
variables of the observable quantities.
Unphysical gauge terms are isolated
and removed by expressing observable quantities in terms of gauge-invariant
variables defined in Appendix~\ref{app:flrw}.

We parametrize the photon geodesic~$x^a(\oo)$ in terms of the affine 
parameter~$\oo$, and its wavevector is
\beeq
k^a(\oo)={dx^a\over d\oo}=\left[~{\bar\nu\over a}~(1+\dnu),~
-{\bar\nu\over a}~(e^\alpha+\dea^\alpha)\right]~,
\label{aeq:wave}
\eneq
where $\bar\nu$ and $e^\alpha$ represent the photon frequency and the photon
propagation direction measured from the observer in the homogeneous universe.
The null condition ($k^ak_a=0$) for the wavevector
constrains the zeroth order propagation direction $e^\alpha e_\alpha=1$, 
and the geodesic equation ($k^bk^a_{~;b}=0$) yields
$\bar\nu\propto1/a$ and $e'^\alpha=e^\beta e^\alpha_{~|\beta}$~.

The dimensionless perturbations to the wavevector $k^a(\oo)$ are 
represented as $\dnu$ and $\dea^\alpha$ for each component
of Eq.~(\ref{aeq:wave}), and for the coordinate transformation in
Eq.~(\ref{aeq:gt}) these perturbation components of the wavevector transform as
\bear
\widetilde{\dnu}&=&\dnu+{d\over d\cc}~T+2\HH T~,  \\
\widetilde{\dea}^\alpha&=&\dea^\alpha+2\HH Te^\alpha-Te'^\alpha-{d\over d\cc}
~\mathcal{L}^\alpha-e^\alpha_{~,\beta}~\mathcal{L}^\beta~,    \nonumber
\enar
where $d/d\cc=\partial_\tau-e^\alpha\partial_\alpha$ and it is related to
the zeroth order photon path $d/d\oo=k^a\partial_a=(\bar\nu/a)(d/d\cc)$~.
Based on the gauge transformation properties, we define two gauge-invariant
variables for the wavevector as
\bear
\dnug&=&\dnu+2H\chi+{d\over d\cc}\left({\chi\over a}\right)~,   \\
\deag&=&\dea^\alpha+2H\chi ~e^\alpha-\left({\chi\over a}\right)e'^\alpha
-{d\over d\cc}~\CCG^\alpha-e^\alpha_{~,\beta}~\CCG^\beta~,\nonumber
\nonumber
\enar
where $\CCG^\alpha=\gamma^{,\alpha}+C^\alpha$ is a pure gauge term,
transforming as $\tilde\CCG^\alpha=\CCG^\alpha-\mathcal{L}^\alpha$.

The dimensionless perturbations $\dnu$ and $\dea^\alpha$ to the wavevector
are also subject to the null condition and satisfy the geodesic equation.
In terms of the gauge-invariant variables, the null condition is
\beeq
e_\alpha ~\deag=\dnug+\ax-\px-\pv_\alpha ~e^\alpha-C_{\alpha\beta}~
e^\alpha e^\beta~,
\label{aeq:null}
\eneq
and the temporal and spatial components of the geodesic equation are
\beeq
{d\over d\cc}~\left(\dnug+2~\ax\right)=(\ax-\px)'-\left(\pv_{\alpha|\beta}
+C'_{\alpha\beta}\right)e^\alpha e^\beta~,
\label{aeq:temp}
\eneq
and
\bear
\label{aeq:sp}
&&\hspace{-15pt}
\left(\deag+2~\px ~e^\alpha+\pv^\alpha+2~C^\alpha_\beta e^\beta
\right)'\\
&&-e^\gamma\left(\deag+2~\px~ e^\alpha+\pv^\alpha+2~C^\alpha_\beta
e^\beta\right)_{|\gamma} \nonumber \\
&&\hspace{-15pt}
=\dea^\beta_\chi e^\alpha_{~|\beta}-\dnug e'^\alpha
+(\ax-\px)^{|\alpha}-\pv_{\beta}{^{|\alpha}}e^\beta-C_{\beta\gamma}
{^{|\alpha}}e^\beta e^\gamma~, \nonumber
\enar
respectively. Fictitious gauge freedoms in $\dnu$ and $\dea^\alpha$ are
completely removed, and Eqs.~(\ref{aeq:null}), (\ref{aeq:temp}), 
and~(\ref{aeq:sp}) are manifestly gauge-invariant. 

As our observable quantities are obtained by measuring photons from galaxies,
we are mainly interested in perturbations along the photon geodesic given
observed redshift~$\zz$ and observed angle~$\nn$. Therefore,
when we consider a coordinate transformation with the
observable quantities fixed,
the affine parameter~$\lambda$ is also affected by the coordinate 
transformation in Eq.~(\ref{aeq:gt}), i.e.,
\beeq
\tilde x^a(\tilde\cc)=x^a(\cc)+\xi^a(\cc)~,
\eneq
and the deviation of the photon geodesic is
\beeq
\widetilde{\delta x^a}=\delta x^a+\xi^a-\bar k^a~{a\over\bar\nu}~\delta\cc~,
\eneq
where $\delta\cc=\tilde\cc-\cc$ and $\widetilde{d\cc}=d\cc~(1-2\HH T)$.
The time lapse and the spatial shift are
\bear
\label{aeq:dto}
\widetilde{\dT_o}&=&\dT_o+T_o~, \\
\widetilde{\delta x_o^\alpha}&=&\delta x_o^\alpha+L_o^\alpha~, \nonumber
\enar
at the origin $\cc=\cc_o$ and
\bear
\label{aeq:dts}
\widetilde{\dT_s}&=&\dT_s+T_s-\delta\cc_s~, \\
\widetilde{\delta x_s^\alpha}&=&\delta x_s^\alpha+L_s^\alpha+e^\alpha\delta\cc_s~, 
\nonumber
\enar
at the source $\cc=\cc_s$, where the subscripts represent that the
quantities are evaluated at the origin $x^a(\cc_o)$ or the source position
$x^a(\cc_s)$.

A comoving observer with the four velocity $u^a=[(1-\alpha)/a,~v^\alpha/a]$
measures the redshift parameter~$\zz$
 from galaxies and it is related to the photon wavevector as
\bear
\label{aeq:zobs}
1+\zz&=&{[k^a(\cc_s)u_a(\cc_s)]\over[k^a(\cc_o)u_a(\cc_o)]}\equiv
{1\over a_s}(1+\dz) \\
&=&{1\over a_s}\bigg\{1+\HH_o\dT_o+\bigg[\dnug+\ax+V-H\chi\bigg]^s_o\bigg\}
\nonumber \\
&=&{1\over a_s}\bigg\{1-H\chi+(H_o\chi_o+\HH_o\dT_o)+
\bigg[V-\ax\bigg]^s_o \nonumber \\
&-&\int_0^{\rs}dr\bigg[(\ax-\px)'-(\pv_{\alpha|\beta}
+C'_{\alpha\beta})~e^\alpha e^\beta\bigg]\bigg\}~, \nonumber
\enar
where $\rs$ is the comoving line-of-sight distance to the source,
$V=V_\alpha e^\alpha$ is the line-of-sight velocity, and $a_s=a[\tau(\cc_s)]$.
The redshift parameter~$z_h$ of the source, commonly defined as
$1+z_h=1/a_s$, needs to be distinguished from the
observed redshift~$\zz$, and Eq.~(\ref{aeq:zobs}) defines the lapse term~$\dz$
in the observed redshift. Spurious spatial gauge freedom in~$\dz$ is removed 
and its temporal gauge dependence is isolated $\widetilde{\dz}=\dz+\HH T$.
We define a gauge-invariant variable for the lapse in the observed redshift
as $\dzg=\dz+H\chi$~.

The spatial displacement of the photon path can be obtained by integrating
the spatial components of the geodesic equation in Eq.~(\ref{aeq:sp})
and using Eqs.~(\ref{aeq:dto}) and~(\ref{aeq:dts}).
The radial displacement is
\bear
\label{aeq:dr}
\delta r&=&e_\alpha ~x_s^\alpha-\rs   \\
&=&(\chi_o+\dT_o)-{\dzg\over\HH}+\bigg[e_\alpha
(\dxg^\alpha-\CCG^\alpha)\bigg]_o^s+\int_o^s d\cc~\dnug \nonumber \\
&=&(\chi_o+\dT_o)-{\dzg\over\HH}-\bigg[e_\alpha \CCG^\alpha\bigg]_o^s 
\nonumber \\
&&+\int_0^{\rs}dr~
\big(\ax-\px-\pv_\alpha e^\alpha-C_{\alpha\beta}e^\alpha e^\beta\big)~, \nonumber
\enar
where $(d/d\cc)~\dxg^\alpha=-\deag$. We have used the null condition
for the wavevector in Eq.~(\ref{aeq:null}) and defined the radial displacement
$\delta r$ with respect to the comoving line-of-sight distance $\rs$,
slightly different from the definition used in \cite{YOFIZA09}.
Similarly, the angular displacements are
\bear
\label{aeq:dt}
\rs~\delta\theta&=&\hat\theta_\alpha x^\alpha_s
=\left[\hat\theta_\alpha(\dxg^\alpha-\CCG^\alpha)\right]_o^s \\
&=&\rs~\hat\theta_\alpha(\deag+\pv^\alpha+2C^\alpha_\beta e^\beta)_o
-\left[\hat\theta_\alpha\CCG^\alpha\right]_o^s \nonumber \\
&&-\int_0^{\rs}dr\bigg[\hat\theta_\alpha(\pv^\alpha+2C^\alpha_\beta e^\beta)
\nonumber \\
&&+\left({\rs-r\over r}\right){\partial\over\partial\theta}
\big(\ax-\px-\pv_\alpha e^\alpha-C_{\alpha\beta}e^\alpha e^\beta\big)
\bigg]~ \nonumber
\enar
and
\bear
\label{aeq:dp}
\rs~\delta\phi&=&\hat\phi_\alpha x^\alpha_s
=\left[\hat\phi_\alpha(\dxg^\alpha-\CCG^\alpha)\right]_o^s \\
&=&\rs~\hat\phi_\alpha(\deag+\pv^\alpha+2C^\alpha_\beta e^\beta)_o
-\left[\hat\phi_\alpha\CCG^\alpha\right]_o^s \nonumber \\
&&-\int_0^{\rs}dr\bigg[\hat\phi_\alpha(\pv^\alpha+2C^\alpha_\beta e^\beta)
\nonumber \\
&&+\left({\rs-r\over r\sin\theta}\right){\partial\over\partial\phi}
\big(\ax-\px-\pv_\alpha e^\alpha-C_{\alpha\beta}e^\alpha e^\beta\big)
\bigg]~, \nonumber
\enar
with the orthonormal direction vectors $\hat\theta$ and $\hat\phi$ 
perpendicular to the observed direction~$\Vang$. The gravitational lensing
convergence is then defined as
\bear
\label{aeq:kappa}
\kappa&=&-{1\over2}\left[\left(\cot\theta+{\partial\over\partial\theta}\right)
\delta\theta+{\partial\over\partial\phi}\delta\phi\right] \\
&=&e_\alpha(\deag+\pv^\alpha+2C^\alpha_\beta e^\beta)_o 
-\int_0^{\rs}dr~ {e_\alpha(\pv^\alpha+2C^\alpha_\beta e^\beta)\over\rs}
\nonumber \\
&+&{1\over2\rs}\int_0^{\rs}dr\left[\hat\theta_\alpha{\partial\over\partial\theta}
+{\hat\phi_\alpha\over\sin\theta}{\partial\over\partial\phi}\right]
(\pv^\alpha+2C^\alpha_\beta e^\beta) \nonumber \\
&+&\int_0^{\rs}dr\left({\rs-r\over 2~r \rs}\right)\hat\nabla^2
\big(\ax-\px-\pv_\alpha e^\alpha-C_{\alpha\beta}e^\alpha e^\beta\big) \nonumber \\
&-&{1\over\rs}\bigg[e_\alpha\CCG^\alpha\bigg]^s_o
+{1\over2}\left[\hat\theta_\alpha{\partial\over\partial\theta}+
{\hat\phi_\alpha\over\sin\theta}{\partial\over\partial\phi}\right]
\CCG^\alpha~, \nonumber
\enar
where the Laplacian operator on a unit sphere is
$\hat\nabla^2=(\partial^2/\partial\theta^2)+\cot\theta(\partial/\partial\theta)
+(1/\sin^2\theta)(\partial^2/\partial\phi^2)$.
The presence of the gauge terms $\CCG^\alpha$ in 
Eqs.~(\ref{aeq:dr})-(\ref{aeq:kappa}) explicitly indicates that
the spatial displacements $\delta r$, $\delta\theta$, and $\delta\phi$ 
and the convergence~$\kappa$ are
gauge-dependent, which can be understood as they represent the difference
between the observed source position and the unobservable unlensed source 
position and subsequently the function thereof. For notational convenience,
we define two gauge-invariant quantities: 
$\drg=\delta r+[e_\alpha\CCG^\alpha]^s_o$ and
$\kag=\kappa+[e_\alpha\CCG^\alpha]^s_o/\rs-[\hat\theta_\alpha(\partial
/\partial\theta)+(\hat\phi_\alpha/\sin\theta)(\partial/\partial\phi)](\CCG^\alpha/2)$.

Finally, the volume occupied by the observed galaxies can be obtained
by tracing backward the photon geodesic $x^a(\cc)$ given the observed
redshift~$\zz$ and angle~$\Vang$, and the total number
of galaxies in the observed volume is
\bear
N_\up{tot}&=&\int\sqrt{-g}~\np~\varepsilon_{abcd}~u^d~
{\partial x^a\over\partial \zz}{\partial x^b\over\partial\theta}
{\partial x^c\over\partial\phi}~d\zz~ d\theta~ d\phi~ \nonumber \\
&=&\int \np~
{r^2\sin\theta \over (1+\zz)^3H}~d\zz ~d\theta ~d\phi\times
~\bigg[1+3~\varphi \nonumber \\
&&+\Delta\gamma+v^\alpha e_\alpha   
+3~\dz+2~{\delta r\over r} 
+H{\partial\over\partial \zz}~\delta r \nonumber \\
&&+\left(\cot\theta+{\partial\over\partial\theta}\right)
\delta\theta+{\partial\over\partial\phi}~ \delta\phi
\bigg]~,
\label{aeq:ngal}
\enar
where $\np$ is the physical number density of the source galaxies,
the metric determinant is $\sqrt{-g}=a^4~(1+\alpha+3~\varphi+\Delta\gamma)$, and
$\varepsilon_{abcd}$ is the Levi-Civita symbol.
Equation~(\ref{aeq:ngal}) means that $N_\up{tot}$ is $\np$ times the physical
volume described by the observed redshift~$\zz$ and angle~$\Vang=(\theta,\phi)$,
and it defines the observed galaxy number density $\nobs$ as
\beeq
N_\up{tot}=\int\nobs
{r^2\sin\theta \over (1+\zz)^3H}~d\zz ~d\theta ~d\phi~,
\eneq
providing the relation to the physical number density $\np$.

With the contributions from the volume distortion 
between the observed and the physical, the observed galaxy number density has 
additional contributions from the physical galaxy number density,
since more galaxies are observed along the overdense
region due to the magnification given the threshold $F_\up{thr}$
for the observed flux,
and this source effect depends on the intrinsic luminosity function of the
source galaxy population, $d\np/dL\propto L^{-s}$: 
$\np=\np(L_\up{thr})(1-5~p~\ddL)$, where the inferred luminosity threshold 
for the observed galaxies is
$L_\up{thr}=4\pi\dLf^2(\zz)F_\up{thr}$, 
the luminosity function slope in magnitude
is $p=0.4(s-1)$, and the luminosity distance in a homogeneous and isotropic
universe is $\dLf$. 

Given the observed redshift~$\zz$ and angle~$\Vang$,
the fluctuation $\ddL$ in the luminosity distance
$\dL$, defined as $\dL(\zz,\Vang)=\dLf(\zz)(1+\ddL)$, is
\cite{SASAK87,YOFIZA09}
\bear
\label{aeq:ddl}
\ddL&=&\HH_o\dT_o+\bigg[\dnu+\alpha+(v_\alpha-\BB_\alpha)e^\alpha\bigg]_s \\
&+&{1\over\rs}\left(\dT_s-{\dz\over\HH_s}\right)
-\int_0^{\rs}dr{(\rs-r)~r\over2~\rs}~
\delta(\hat R_{ab}\hat k^a\hat k^b) \nonumber \\
&=&(\HH_o\dT_o+H_o\chi_o)+\left({\chi_o+\dT_o\over \rs}\right)
-\ax+V-{\dzg\over\HH\rs}\nonumber \\
&-&\int_0^{\rs}dr~{r\over \rs}~
\bigg[(\ax-\px)'-(\pv_{\alpha|\beta}+C'_{\alpha\beta})
e^\alpha e^\beta\bigg] \nonumber \\
&+&{1\over \rs}\int_0^{\rs}dr~2\ax-\bigg[H_o\chi_o+H\chi-
{1\over\rs}\int_0^{\rs}dr~2H\chi \nonumber \\
&+&\int_0^{\rs}dr{(\rs-r)~r\over2~\rs}~
\delta(\hat R_{ab}\hat k^a\hat k^b)\bigg]~, \nonumber 
\enar
and it is related to the magnification~$\mu$ as $\mu=1-2~\ddL$.
The last term in Eq.~(\ref{aeq:ddl}) is the
source term for the expansion of the wavevector $\vartheta=\hat k^a_{~;a}$
in the conformally transformed metric $\hat g_{ab}=(\bar\nu/a)g_{ab}$
with the corresponding affine parameter~$\cc$,
\bear
&&\hspace{-10pt}\delta(\hat R_{ab}\hat k^a\hat k^b)=
\Delta(\ax-\px)-\ax{_{,\alpha|\beta}}~e^\alpha e^\beta-2~{d^2\over d\cc^2}~\px
\nonumber \\
&&+(\pv'_{\alpha|\beta}e^\beta-\Delta\pv_\alpha)e^\alpha+
(C''_{\alpha\beta}-\Delta C_{\alpha\beta})e^\alpha e^\beta~,
\enar
where $\hat R_{ab}$ and $\Delta$ are the Ricci tensor and Laplacian operator
of a three-space.

The last ingredient of our formalism is the linear bias approximation that
relates the physical galaxy number density $\np$ to the underlying matter 
distribution $\rho_m$. Observationally, the time slicing is set by the observed
redshift~$\zz$ and this choice is the only physical way in cosmological
observations for defining 
the hypersurface of simultaneity without gauge ambiguity. 
The matter density at the observed redshift~$\zz$ of source galaxies is
$\rho_m=\bar\rho_m(t)(1+\delta_m)=\bar\rho_m(\zz)(1+\DD)$ 
where $\DD=\delta_m-3~\dz$ and it is simply the matter fluctuation
$\DD=\delta_m$ in the uniform-redshift
gauge ($\dz=0$). Therefore, the simplest
linear bias ansatz we adopt is $\np=\bnobs(\zz)(1+b~\DD)$~.

Finally the observed galaxy number density 
can be written in a manifestly gauge-invariant manner as
\bear
\label{aeq:dobs}
\hspace{-20pt}
\nobs=\bnobs(\zz)\bigg[1&+&b~\DD+\ax+2~\px+V \\
&-&C_{\alpha\beta}~e^\alpha e^\beta+3~\dzg+2~{\drg\over\rs} \nonumber \\
&-&H{\partial\over\partial\zz}\left({\dzg\over\HH}\right)
-5p~\ddL-2~\kag\bigg]~, \nonumber
\enar
and this completes the derivation of Eq.~(\ref{eq:full}). 
In Eq.~(\ref{aeq:dobs})
 $\bnobs(\zz)$ is the mean galaxy number density
in a homogeneous and isotropic universe, in a close analogy to $\bar\rho_m$,
though we cannot predict a priori its functional form. Lastly,
it is worth emphasizing that
Eq.~(\ref{aeq:dobs}) is valid to the linear order only
in metric perturbations as it involves no linearization in the
matter density fluctuation~$\delta_m$, and hence it is valid in principle
deep in the Newtonian limit, where the validity of the linear 
bias approximation is, however, highly in suspect.

\end{document}